\documentclass[journal]{IEEEtran}
\IEEEoverridecommandlockouts
\usepackage{cite}
\usepackage{algorithm,amsbsy,amsmath,amssymb,epsfig,bbm,mathrsfs,multirow,amsthm,amsfonts}
\usepackage{array}
\usepackage{graphicx}
\usepackage{graphics}
\usepackage{subfigure}
\usepackage{epstopdf}
\usepackage{color}
\usepackage{soul}
\usepackage{bbm}
\usepackage{algorithm}
\usepackage[noend]{algpseudocode}
\usepackage{mathtools}
\usepackage{pdfpages}
\usepackage{textcomp}
\usepackage{xcolor}

\algrenewcommand\algorithmicforall{\textbf{foreach}}
\algrenewcommand\algorithmicindent{.8em}

\def\BibTeX{{\rm B\kern-.05em{\sc i\kern-.025em b}\kern-.08em
    T\kern-.1667em\lower.7ex\hbox{E}\kern-.125emX}}

\begin{document}

\title{Trustworthy GenAI over 6G: Integrated
Applications and Security Frameworks}

\author{ {Bui Duc Son},~\IEEEmembership{Graduate Student Member,~IEEE}, Trinh Van Chien,~\IEEEmembership{Member,~IEEE},\\ and {Dong In Kim},~\IEEEmembership{Life Fellow,~IEEE}
 \thanks{This research was supported in part by the MSIT (Ministry of Science and ICT), Korea, under the ICT Creative Consilience program (IITP-2020-0-01821) supervised by the IITP (Institute for ICT Planning \& Evaluation). (Corresponding author: Dong In Kim)}
\thanks{Bui Duc Son and Dong In Kim are with the Department of Electrical and Computer Engineering, Sungkyunkwan University, Suwon 16419, South Korea. Emails: buiducson@skku.edu and dongin@skku.edu.}
\thanks{ Trinh Van Chien is with the School of Information and Communications Technology, Hanoi University of Science and Technology, Hanoi 100000, Vietnam. Emails:  chientv@soict.hust.edu.vn.}
}

\maketitle

\begin{abstract}
\textcolor{black}{The integration of generative artificial intelligence (GenAI) into 6G networks promises substantial performance gains while simultaneously exposing novel security vulnerabilities rooted in multimodal data processing and autonomous reasoning. This article presents a unified perspective on cross-domain vulnerabilities that arise across integrated sensing and communication (ISAC), federated learning (FL), digital twins (DTs), diffusion models (DMs), and large telecommunication models (LTMs). We highlight emerging adversarial agents such as compromised DTs and LTMs that can manipulate both the physical and cognitive layers of 6G systems. To address these risks, we propose an adaptive evolutionary defense (AED) concept that continuously co-evolves with attacks through GenAI–driven simulation and feedback, combining physical-layer protection, secure learning pipelines, and cognitive-layer resilience. A case study using an LLM-based port prediction model for fluid-antenna systems demonstrates the susceptibility of GenAI modules to adversarial perturbations and the effectiveness of the proposed defense concept. Finally, we summarize open challenges and future research directions toward building trustworthy, quantum-resilient, and adaptive GenAI–enabled 6G networks.}

\end{abstract}

\begin{IEEEkeywords}
GenAI, 6G networks, adversarial attacks, integrated sensing and communication (ISAC), federated learning, cross-domain security.
\end{IEEEkeywords}

\section{Introduction}

\textcolor{black}{Generative artificial intelligence (GenAI) is emerging as a key enabler for intelligent mobile services. Applications such as digital twins (DTs) allow real-time simulation and interaction with the physical world, while diffusion models (DMs) and large language models (LLMs) provide powerful reasoning and data generation capabilities \cite{10599121}. In parallel, sixth-generation (6G) networks are envisioned as intelligent infrastructures that will support integrated sensing and communication (ISAC), edge/cloud offloading, and distributed learning, creating perception–communication–computation loops essential for autonomous vehicles, smart factories, and immersive services \cite{xu2024large}. The convergence of 6G and GenAI is mutually reinforcing. The former offers ultra-low latency and dense data streams required for training and deploying complex GenAI models. In return, the latter contributes adaptive optimization and intelligent allocation of wireless resources. This collaboration promises a new class of autonomous, scalable, and interactive wireless ecosystems. Nevertheless, such deep integration significantly expands the attack surface. Unlike prior generations, the boundary between 6G and GenAI has blurred: perturbations at the physical layer can directly corrupt GenAI perception, while poisoned models or malicious prompts can propagate downward to disrupt connectivity \cite{zhao2024generative}. For example, over-the-air malicious transmissions may be sufficient to mislead model decisions, while illegal devices in federated learning (FL) can poison global models and degrade network resource allocation \cite{yazdinejad2024robust}. Additionally, through data poisoning and modification of the training objective, attackers are able to introduce backdoor triggers into diffusion models \cite{truong2025attacks}. These dynamics expose vulnerabilities that conventional approaches, which address wireless and GenAI security separately, cannot resolve.}

\textcolor{black}{Existing works have primarily examined security challenges in either the wireless physical layer or adversarial machine learning, but rarely in a unified manner. For example, physical layer defenses against eavesdropping and jamming are well-studied \cite{zhao2024generative}, while research into adversarial learning has focused on the resilience of deep neural networks and FL models \cite{10460991}. Recent surveys also discuss ISAC security and 6G, and there is emerging work on trustworthy DTs \cite{chai2024generative}. However, these efforts largely address each technology separately, without analyzing how vulnerabilities can propagate across ISAC, FL, DTs, DMs, and LTMs \cite{11097898} when they are jointly deployed. Moreover, the potential of adversarial DTs or malicious LTMs acting as active threat agents remains overlooked, as well as empirical validation of such cross-domain risks under realistic 6G AI-native architectures is still limited. By carefully considering these concerns, it is crucial to investigate a cross-domain attack and defense to ensure the sustainability of future wireless communication systems and beyond. This framework must be adaptive and founded on the principle of co-design between the wireless and AI layers. Hence, our main contributions are summarized as follows:}
\begin{itemize}
    \item \textcolor{black}{We present a novel integrated security architecture spanning ISAC, FL, DTs, and LTMs, explicitly showing how local vulnerabilities can propagate across layers.}
    \item \textcolor{black}{We introduce adversarial DTs and malicious LTMs as active threat agents, extending beyond prior work that viewed them as passive targets, bringing novel adversarial threat models.}
    \item \textcolor{black}{We introduce an adaptive evolutionary defense (AED) concept that enables the system to co-evolve with adversaries through continuous policy generation, GenAI–based attack simulation, and real-time feedback coordination, providing long-term robustness under dynamic 6G environments.}
    \item \textcolor{black}{Using Port-LLM as a case study, we demonstrate that LLM-based telecom modules are highly vulnerable to malicious input manipulations, highlighting the urgency of cross-layer defenses.}
    \item \textcolor{black}{Finally, we discuss key challenges and open issues for securing GenAI–enabled 6G networks, including adaptive adversaries, resource constraints, standardization gaps, and the emerging impact of quantum-accelerated attacks and defenses}
\end{itemize}

\section{Core Enabling Technologies for Collaboration between GenAI Agents and 6G Networks}
\begin{figure}[t]
    \centering
    \includegraphics[width=1\linewidth]{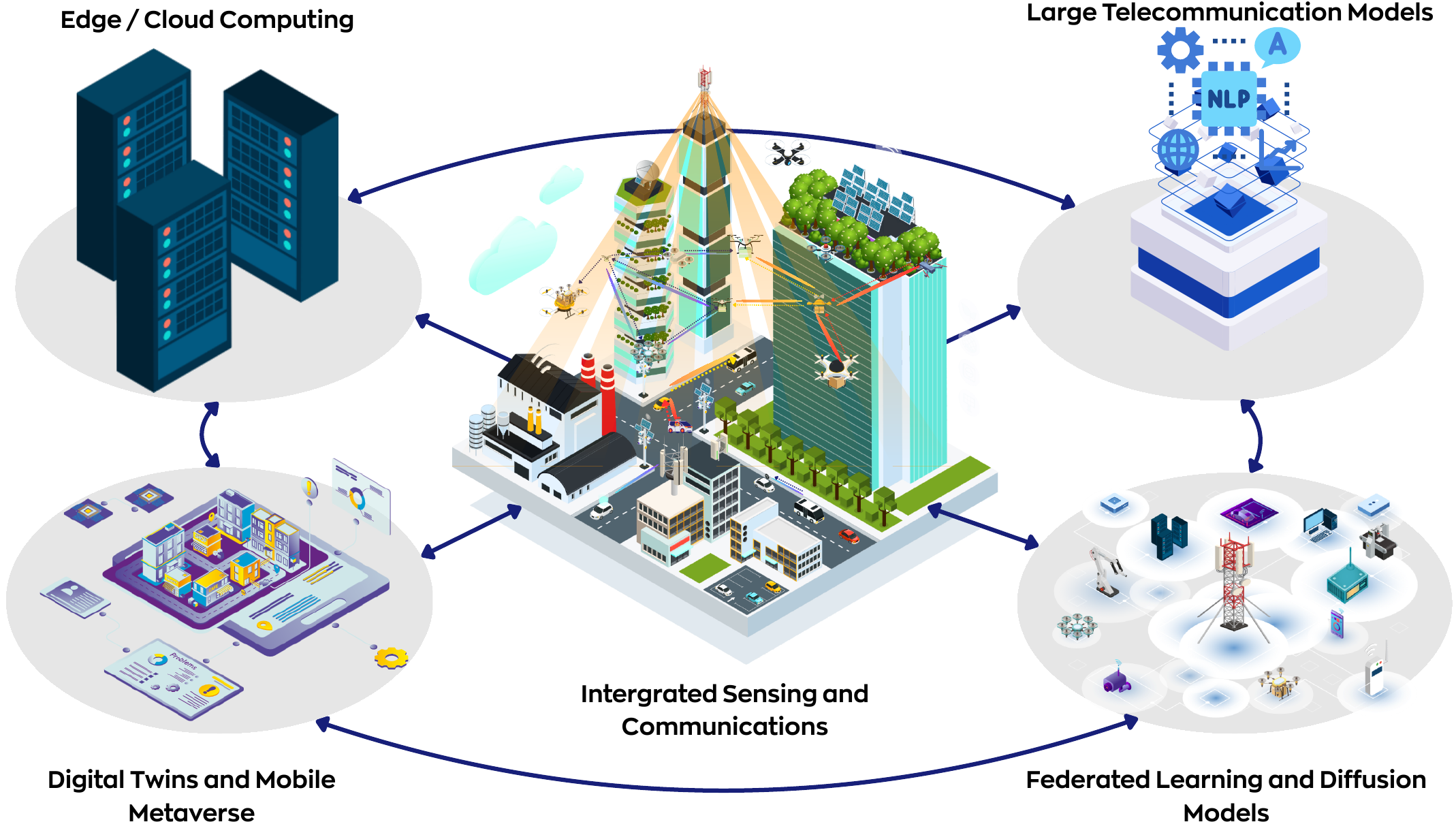}
    \caption{Core enabling technologies supporting GenAI-enabled agents in 6G networks.}
    \label{fig:fivetech}
    \vspace{-0.5cm}
\end{figure}
\textcolor{black}{The integration of GenAI agents into 6G networks builds on several foundational technologies that enable real-time perception, distributed intelligence, and synchronized decision-making. This section highlights five key enablers as illustrated in Fig.~\ref{fig:fivetech} and then briefly discusses their unified role in supporting GenAI-aided 6G networks.}

\subsection{Sensing and Perception with Integrated Sensing and Communication}
\textcolor{black}{ISAC enables sensing and communication on shared radio resources, allowing GenAI agents to perceive their physical surroundings directly through the wireless infrastructure \cite{11062661}. This collaboration allows the collection of real-time multimodal data while maintaining reliable connectivity, supporting applications such as autonomous driving, robotic platforms, and industrial automation. In addition, multimodal fusion is especially critical in these environments, where agents must operate safely under uncertainty \cite{xin2024novel}. However, because ISAC outputs can be directly fed to GenAI inference loops, even minor input manipulations can degrade perception and control.}

\subsection{Federated Learning for Distributed Intelligence}
\textcolor{black}{FL enables distributed, privacy-preserving model training among GenAI agents without requiring raw data exchange \cite{huang2024federated}. Each agent trains locally on device-specific data and transmits only model updates or gradients, reducing communication overhead and protecting sensitive information. Edge servers or cloud nodes periodically aggregate these updates into a global model, making FL suitable for large-scale, bandwidth-limited 6G deployments. Nonetheless, the distributed nature of FL also exposes it to poisoning attacks and adversarial manipulations, which can propagate through the aggregation process. Secure aggregation protocols and anomaly detection are thus critical to maintaining reliable distributed intelligence.}

\subsection{Digital Twins for Synchronization}
\textcolor{black}{DTs serve as dynamic, virtual replicas of physical environments, continuously updated with ISAC-derived sensory data and other sources \cite{chai2024generative}. They can provide a shared cognitive backbone that allows GenAI agents to validate observations, anticipate dynamics, and coordinate actions across heterogeneous systems. In multi-agent scenarios such as unmanned aerial vehicle (UAV) swarms or industrial robotics, DTs help maintain synchronized decision-making and enhance resilience. However, if manipulated, altered  DT states can mislead perception and coordination, introducing cascading failures. Ensuring trustworthiness verification and anomaly detection within DT frameworks is therefore vital.}

\subsection{Diffusion Models for Data Generation and Environment Modeling}
\textcolor{black}{DMs are emerging as powerful generative tools within 6G-enabled AI pipelines \cite{zhang2024denoising}. Beyond synthetic data generation, DMs can capture the stochastic properties of wireless environments, such as mmWave and terahertz channels, enabling robust simulations for beamforming, resource allocation, and mobility management. At the network edge, conditional DMs also support on-device data augmentation, extending local datasets to improve FL performance while preserving privacy. Nevertheless, reliance on synthetic data introduces a new vulnerability: compromised generative models may yield poisoned or misleading samples, undermining system reliability. Continuous validation and monitoring of diffusion-based pipelines are therefore required.}

\subsection{Large Telecommunication Models for Domain-Specific Cognition}
\textcolor{black}{TMs embed domain-specific telecom knowledge into reasoning, optimization, and decision-making across the 6G network \cite{11097898}. Trained on protocol specifications, performance logs, user behavior data, and multimodal sensing inputs, LTMs support intelligent scheduling, anomaly detection, and cross-layer optimization. Depending on latency and resource constraints, LTMs can be deployed as lightweight models on devices, mid-sized models at the edge, or large-scale engines in the cloud. While LTMs offer explainability and domain grounding, they also face attack surfaces: adversarial prompts, poisoned updates, and manipulated data can misdirect decisions and degrade the quality of service. Safeguards such as prompt filtering, anomaly detection, and integrity verification are essential as LTMs become the cognitive engines of GenAI-aided 6G networks.}

\section{Attack Strategies for Individual Components}
\begin{figure}[t]
    \centering
    \includegraphics[width=0.7\linewidth]{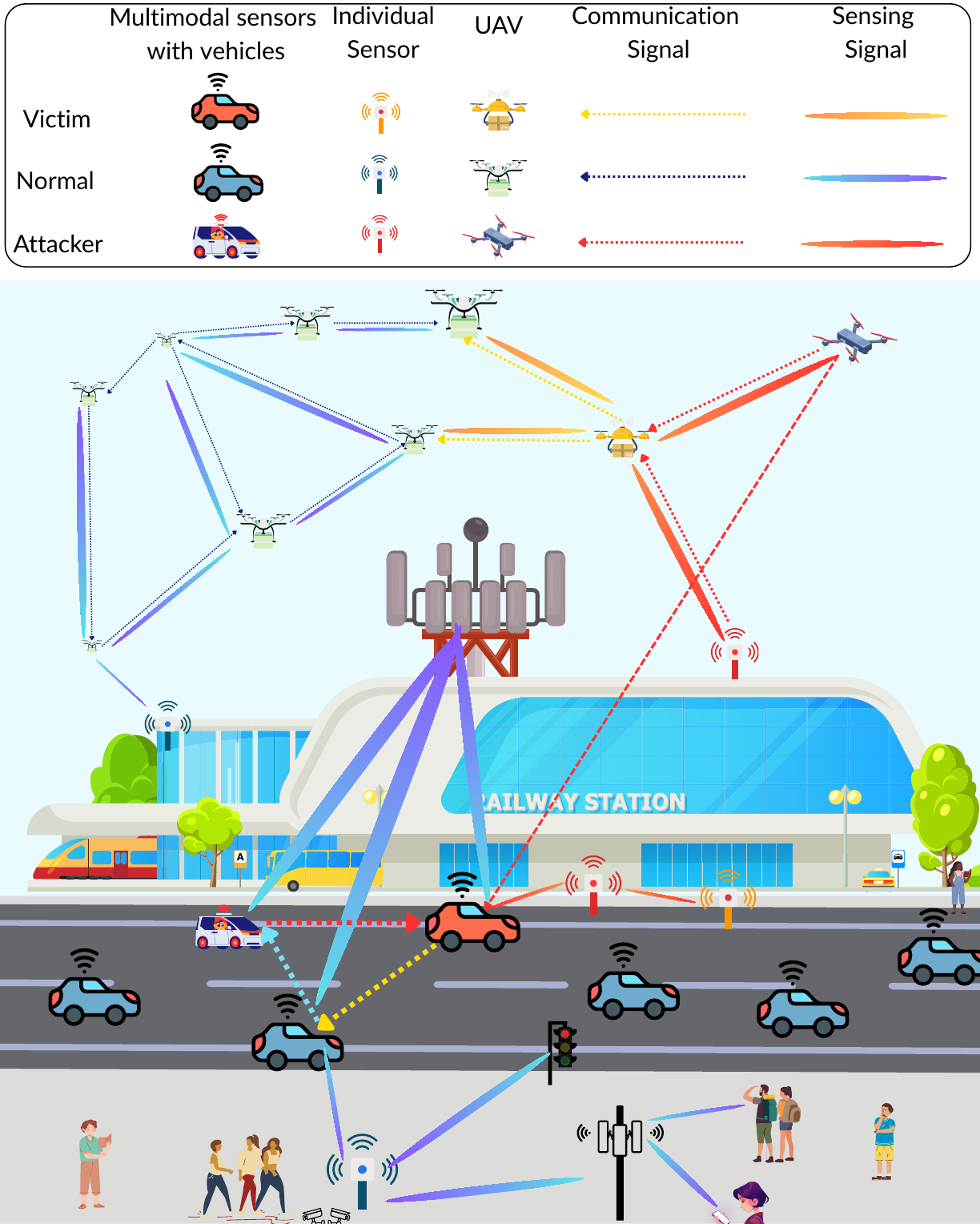}
    \caption{Cross-domain attack paths targeting sensing, learning, and reasoning components in a 6G-enabled environment.}
    \label{fig:ISACATTACK}
    \vspace{-0.5cm}
\end{figure}

\textcolor{black}{The integration of multimodal sensing, distributed learning, generative modeling, and cognitive reasoning in GenAI-assisted 6G networks significantly broadens the attack surface. Adversaries can simultaneously exploit vulnerabilities at the physical layer, compromise learning processes, and manipulate reasoning mechanisms. This section explores the potential weaknesses of the core technologies that enable collaboration between GenAI agents and 6G networks.}

\subsection{Attacks on Sensing and Communication Signals}
\textcolor{black}{ISAC-enabled perception makes GenAI agents vulnerable to harmful changes introduced before or during digital preprocessing. Because RF signals are broadcast and sensors operate in open environments, attackers can perturb multimodal inputs over-the-air to bypass conventional defenses. One approach is to inject specially crafted waveforms, where small disturbances are created using algorithms like the Fast Gradient Sign Method (FGSM), Projected Gradient Descent (PGD), or the Carlini \& Wagner (C\&W) attack, applied in either white-box or transfer-based scenarios. These waveforms, designed to distort sensed features while remaining power-efficient, can be transmitted through IoT devices or compromised nodes to bias perception and decision-making in real time. Another method involves replay and signal forgery attacks, which exploit the temporal consistency assumptions of ISAC modules. Adversaries can record legitimate signals and retransmit outdated or modified versions, desynchronizing system perception from the real physical environment. Such attacks are typically black-box, making them both practical and difficult to detect. For instance, as shown in Fig.~\ref{fig:ISACATTACK}, in a UAV swarm, a malicious UAV may inject forged signals to mislead others, triggering cascading errors across the formation. Similarly, in sensor networks, adversarial nodes may feed outdated data to misguide autonomous vehicles, potentially leading to unsafe decisions.}

\subsection{Attacks on Federated Learning and Diffusion Models}
\textcolor{black}{In GenAI-assisted 6G–ISAC systems, FL aggregates updates over unreliable wireless links, while DMs generate or augment data at the edge. Their decentralized and partly untrusted nature introduces multiple attack vectors. To damage FL, adversaries may perform label-flipping attacks, where incorrect labels shift the global decision boundary. This is especially harmful under non-IID data distributions, where it is difficult to distinguish abnormal gradients from natural variability. Model poisoning attacks exploit robust aggregators by crafting gradients that stay within norm bounds or inner-product metrics, thereby bypassing detection and even embedding backdoors that activate under specific triggers. DMs face distinct threats, as perturbations to noise schedules, score networks, or conditioning vectors can generate poisoned synthetic data that degrades subsequent training. Moreover, data-free and gradient-leakage attacks exploit the exposure of gradients in FL. Using gradient inversion or side-channel analysis, adversaries can reconstruct surrogate inputs or craft malicious updates without direct access to the original ISAC data. These attack vectors can reinforce one another: compromised DMs can continuously feed poisoned data into FL, while corrupted FL updates degrade future DM training. Such self-reinforcing loops erode both data quality and model trustworthiness, underscoring the need for end-to-end defenses across the wireless communications ecosystem.}

\subsection{Attacks on Large Telecommunication Models}
\textcolor{black}{LTMs serve as the cognitive layer in GenAI-assisted 6G systems, integrating network control signals, sensing feedback, and user prompts into unified decision pipelines. This role exposes LTMs to both input-time manipulation and training-time poisoning. At the input level, prompt injection and jailbreak attacks exploit the context-driven nature of LTMs. Since safeguards often rely on protocol or parameter constraints without cryptographic protection, malicious instructions embedded in sensor metadata, control messages, or DT states can override safety policies. Once processed, the LTM may leak sensitive data or disable critical protections. At the training level, data poisoning attacks are a major threat in continual or FL settings. Even a small fraction of corrupted data, such as mislabeled adversarial gradients, can shift model boundaries or implant backdoors that activate under specific tokens or signals. Finally, reasoning manipulation attacks alter the inference process itself. By tampering with prompt templates, control instructions, or DT state logs, adversaries can redirect LTM reasoning without triggering anomalies in loss metrics or gradient statistics. Although outputs may appear valid, the model may in fact be executing malicious objectives, compromising network reliability and trustworthiness.}
\begin{figure*}[ht]
    \centering
    \includegraphics[width=0.8\linewidth]{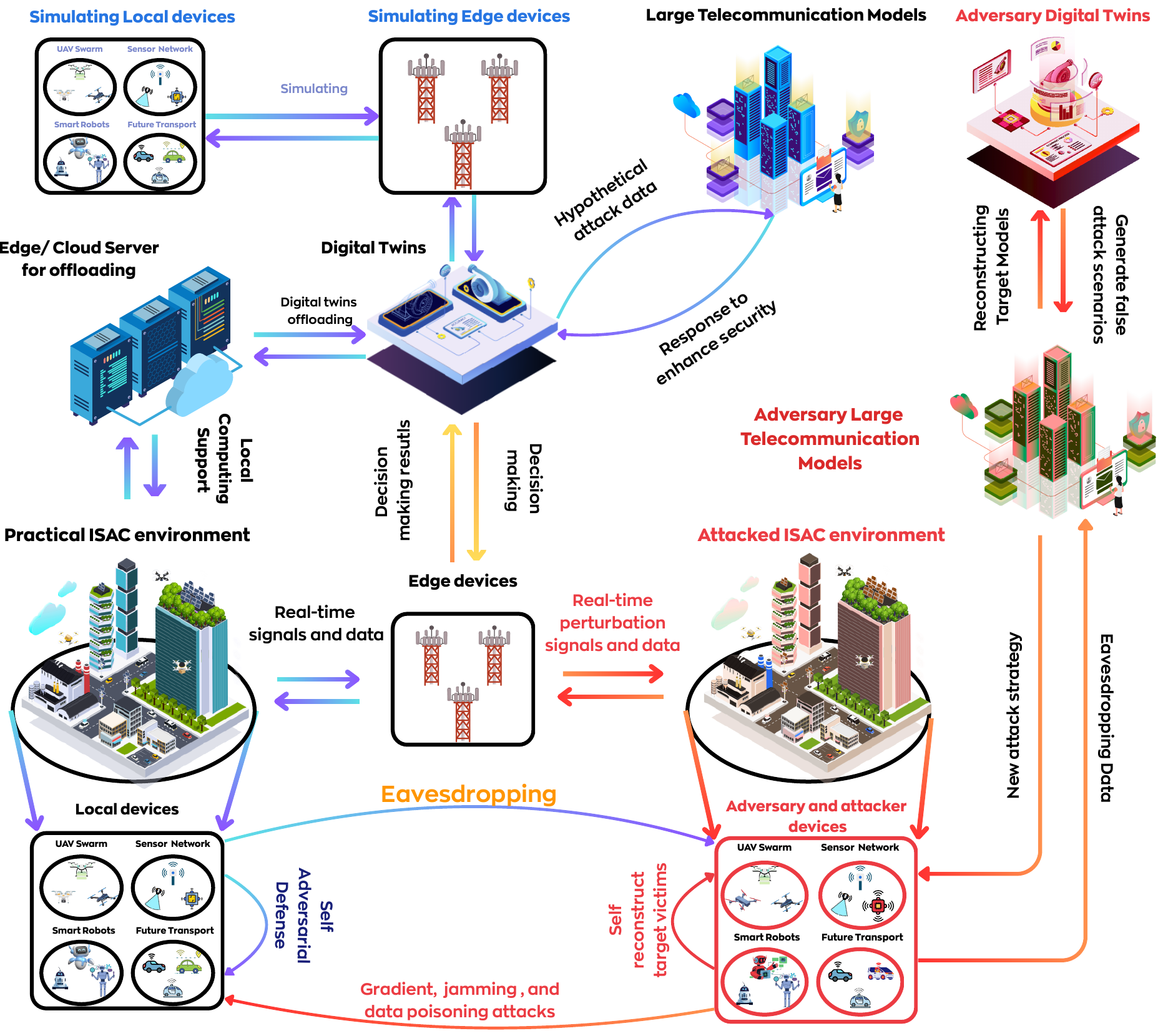}
    \caption{Unified cross-layer framework illustrating how adversarial behaviors can propagate across ISAC, FL, DTs, and LTMs.}
    \label{fig:attack_overview}
    \vspace{-0.5cm}
\end{figure*}

\subsection{Summary of Attack Strategies}
\textcolor{black}{The attacks described across sensing, learning, and reasoning layers reveal a multi-domain and mutually reinforcing threat landscape in GenAI-assisted 6G networks. At the physical layer, adversaries exploit the broadcast nature of wireless signals through perturbations, replay, or forgery to mislead ISAC modules. At the learning layer, FL updates and DMs are corrupted via label flipping, poisoning, or gradient leakage, creating feedback loops that degrade model integrity over time. Finally, at the cognitive layer, LTMs are exposed to prompt injection, data poisoning, and reasoning manipulation, undermining decision reliability and network trustworthiness. Collectively, these threats highlight the urgent need for cross-layer security frameworks that account for the interdependencies among physical signals, learning pipelines, and high-level reasoning in 6G systems.}

\section{Cross-Domain Attack for GenAI-Enabled 6G Networks}

\textcolor{black}{This section introduces a unified framework for GenAI-enabled 6G networks. As illustrated in Fig.~\ref{fig:attack_overview}, the architecture integrates two core components, DTs and LTMs. For simplicity, we define the frameworks consisting of three layers:}
\begin{itemize}
    \item \textbf{Local (physical) layer}: Focused on ISAC operations.
    \item \textbf{Edge/cloud layer}: Responsible for offloading, synchronization, and virtualization.
    \item \textbf{Cognitive layer}: Supporting reasoning and high-level decision-making.
\end{itemize}
\textcolor{black}{The framework is built upon a tightly coupled loop of perception, learning, generation, and reasoning that operates seamlessly across devices, edge servers, and cloud infrastructures. When adversaries are present, they can also exploit DTs and LTMs to generate sophisticated attacks, resulting in a dynamic competition between attack strategies and defense mechanisms. }
\subsection{Standard Operation Workflow}
\textcolor{black}{At the local layer, devices such as UAVs, distributed sensors, and industrial robots continuously produce real-time multimodal data. This data is processed at nearby edge servers to reduce latency and computational burden, and selectively offloaded to cloud servers for large-scale analytics and long-term knowledge storage.
The DT provides a synchronized virtual replica of the physical environment, enabling predictive modeling, resource optimization, and proactive orchestration of services. Complementing this, the LTM serves as the cognitive engine, fusing multimodal streams, control signals, sensing feedback, and contextual prompts to provide high-level reasoning and task-oriented decision support.
Through this hierarchical and interconnected design, data flows seamlessly across ISAC modules, edge/cloud intelligence, DTs, and LTMs. As a result, autonomous systems such as UAV swarms, connected vehicles, and industrial robots operate with enhanced accuracy, efficiency, and responsiveness, laying the foundation for scalable and resilient next-generation wireless services.}

\subsection{Adversarially Compromised Operation}
\textcolor{black}{The same feedback mechanisms that optimize performance can also create pathways for cascading failures when exploited by adversaries. The attack surface spans all three layers:}
\begin{itemize}
    \item \textbf{Local layer}: Adversaries may inject perturbations, replay outdated information, or poison data, directly distorting ISAC outputs.
    \item \textbf{Edge/cloud layer}: By eavesdropping or impersonating clients, attackers can reconstruct traffic, conduct gradient-based attacks, or selectively jam communication links. These actions desynchronize DT from the physical world, leading to misleading states.
    \item \textbf{Cognitive layer}: Malicious prompts or tainted training pipelines induce biased reasoning processes, ultimately compromising the reliability of cognitive decisions.
\end{itemize}
\textcolor{black}{Importantly, the attacker’s DT and LTM can be used to design and test attack scenarios, guess how defenses will respond, and figure out ways to get around them. The combined impact of signal distortion, DT desynchronization, and poisoned reasoning propagates through the coupled perception-learning-generation-reasoning loop. Ultimately, this produces corrupted actuation decisions, compromising the safety, autonomy, and trustworthiness of UAV swarms, connected vehicles, industrial robots, and other intelligent 6G-enabled services.}

\section{Adaptive Evolutionary Defense Concept}
\begin{figure}[t]
    \centering
    \includegraphics[width=1\linewidth]{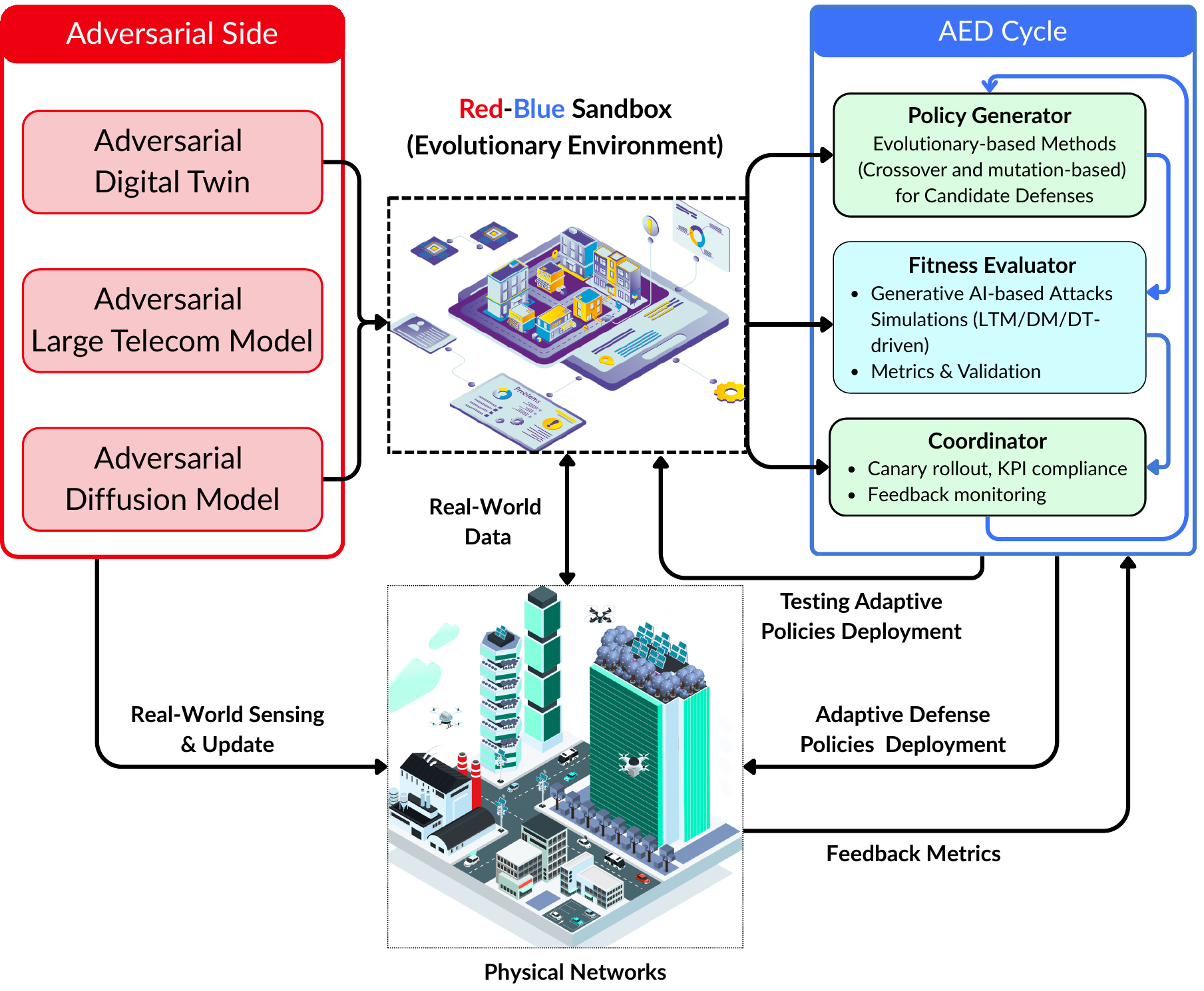}
    \caption{Overview of the proposed AED concept.}
    \label{fig:aed}
    \vspace{-0.5cm}
\end{figure}

\textcolor{black}{In modern wireless and AI-enabled systems, adversaries evolve rapidly to bypass defenses, raising a key question: \textbf{When is a system still secure?} Thus, we define security through operational performance bounds: the upper bound reflects ideal performance under deployed defenses, while the lower bound sets the minimum acceptable level ensuring service continuity. \textbf{A system is secure if its performance under any attack remains above this lower bound. This shifts the goal from eliminating all threats, which is impossible, to guaranteeing resilience.} To achieve this, we propose the AED concept, as illustrated in Fig.~\ref{fig:aed}, which leverages co-evolution to dynamically adapt defenses and maintain robustness against evolving adversaries in GenAI-assisted 6G networks.}

\subsection{Evolution environments}

\textcolor{black}{In the black-box setting, initialization-based attacks are largely ineffective, forcing adversaries to learn and adapt their strategies, which can be viewed as an evolutionary process. Adversaries exploit GenAI to discover vulnerabilities through techniques such as reconstruction and eavesdropping, targeting both training and inference phases. Thus, the system must not only defend but also continuously learn to mitigate attacks in real time. Since neither defense nor attack method is perfect, the only viable strategy is to enable adaptive, self-evolving defense mechanisms that preserve robustness in dynamic environments. This co-evolutionary process between the attacker and defender defines what we term the \textbf{evolution environments}.} 

\subsection{Adaptive Evolutionary Defense: Goals and Key Components}

\textcolor{black}{The primary goal of the AED concept is not to completely eliminate attacks, as that is practically infeasible. Instead, it aims to ensure that the system's robustness consistently remains above a defined lower-bound threshold. This approach guarantees that, regardless of the intensity or sophistication of an attack, the system's degradation remains within an acceptable range. Therefore, AED acts as a proactive and self-adjusting shield, providing long-term resilience in highly adversarial and uncertain 6G environments. To realize this objective, AED operates as a continuous evolutionary cycle with three interconnected components:}
\begin{itemize}
    \item \textbf{Policy Generator (PG):} \textcolor{black}{Drawing on evolutionary learning principles, the PG self-directs the adaptive search and refinement of defense strategies across layers. LTM provides structural priors that constrain the PG’s search space, ensuring that evolved strategies remain physically feasible and deployable. To prevent convergence to an overly specialized and evolution attack-sensitive defense strategy, the PG periodically injects policy diversity by perturbing and recombining previously robust solutions. Moreover, to cope with non-stationary wireless conditions, the PG is guided by a DM-based dynamic DT, which forecasts likely future propagation environment and attack patterns. This allows the PG to evolve defenses ahead of the adversary, instead of reacting only after degradation occurs.}
    
    \item \textbf{GenAI-based Fitness Evaluator (FE):} \textcolor{black}{The FE serves as a high-fidelity red-team engine that evaluates candidate defense policies under diverse adversarial conditions, including gradient-based perturbation, pilot contamination, selective jamming, and data poisoning. A physics-aware validation layer, informed by LTM-derived knowledge of wireless propagation and protocol behavior, ensures that simulated attacks remain realistic and physically consistent. In addition, the FE incorporates agentic self-assessment, enabling it to detect when its learned attack models become outdated and autonomously request updated environmental simulations from the DM-based dynamic DT. This ensures that policy evaluation remains aligned with real, evolving threat landscapes.} 

    \item \textbf{Coordinator (CO):} \textcolor{black}{Acting as the control hub between simulation and deployment, the CO manages the safe rollout of the most resilient defense policies across the system. The CO performs phased deployment and continuously monitors key performance indicators (KPIs), reverting to validated baseline defense policies when system performance approaches the required robustness threshold. To enable continuous adaptation, the CO utilizes a multi-agent reinforcement learning controller that fine-tunes defense behavior in real-time based on feedback, while LTM priors regularize adaptation to prevent unstable or infeasible policy shifts. Critically, the CO integrates the agentic AI self-learning mechanism, which autonomously identifies when existing defense knowledge becomes insufficient, formulates new adaptation goals, and initiates renewed PG–FE evolution cycles. This elevates AED from a reactive scheme to a continuously self-improving, self-motivated defense framework.}

\end{itemize}
\textcolor{black}{This loop repeats continuously, creating a self-improving security system that learns from each attack and gradually enhances the network's security. The diffusion-based environment forecasting enables AED to anticipate shifts in channel and adversarial behavior, rather than reacting only after degradation occurs. Meanwhile, the reinforcement learning–driven online policy refinement maintains stable robustness as conditions evolve, ensuring that the co-evolving defense remains effective even in dynamic and non-stationary 6G environments. However, in the case of an unpredictable or entirely new attack pattern, even when supported by DM-based forecasting, LTM-guided priors, and DT-driven environment modeling, AED does not attempt to extrapolate beyond its confidence. Instead, the CO detects the increased uncertainty and shifts the system into a robust fallback mode, where proven and stable defense strategies are applied to sustain acceptable performance while the new adversarial behavior is observed and gradually learned.}

\section{Case Study: Adversarial Attack against Large Language Model for Wireless Communication}

\textcolor{black}{In this section, we evaluate the Port-LLM model \cite{11203229}, a large language model designed for fluid-antenna port prediction, under an adversarial setting where the attacker gradually adapts its strategy over time. This setup allows us to observe how the model’s performance degrades as the adversary becomes more sophisticated, and how the proposed AED mechanism responds to restore and maintain performance stability.}


\subsection{Adversarial Strategy and Evaluation Protocol}
The adversary employs gradient-based adversarial attacks to perturb the input channel samples, thereby reducing the port-prediction accuracy of Port-LLM. The attacker adjusts its perturbations by observing the model’s loss, allowing the perturbation direction to evolve throughout training. The adversary’s objective is to push the model’s accuracy below a predefined threshold.

Model evaluation is performed at the end of each training epoch. During testing, the adversary generates gradient-based perturbations on the fly, enabling us to assess robustness under real-time attack conditions. This protocol provides a consistent measurement of how Port-LLM responds to increasingly challenging adversarial inputs once training converges.

\begin{figure}[t]
	\centering
    \begin{minipage}[t]{0.3\textwidth}
	\includegraphics[width=1\textwidth]{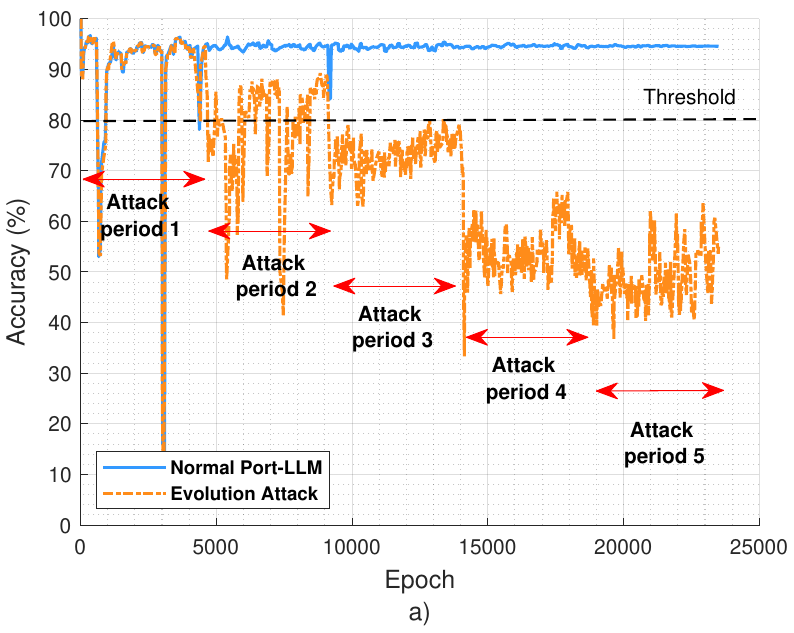}
    \end{minipage}
    \centering
    \begin{minipage}[t]{0.3\textwidth}
	\includegraphics[width=1\textwidth]{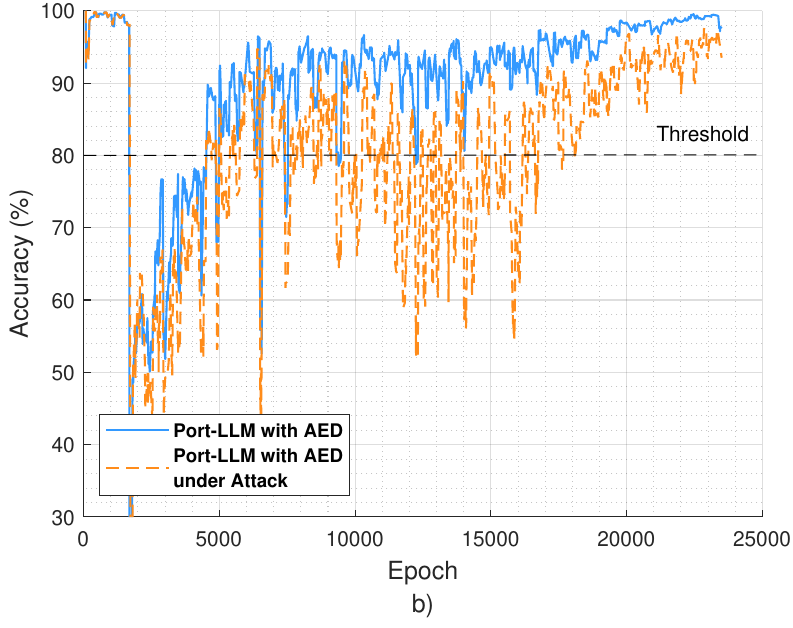}
    \end{minipage}
\caption{Performance of Port-LLM under adversarial attack and with the proposed AED: (a) Performance of Port-LLM under evolving adversarial attacks. (b) Improved stability when the AED is applied.}
\vspace{-0.5cm}
\label{fig:PortLLM_AED}
\label{losss}
\end{figure}

\subsection{Integration of AED Concept into the Port-LLM Case Study}
\textcolor{black}{To validate AED, we integrate it into the Port-LLM training framework. In this setup, the adversary applies evolving gradient-based attacks, while candidate defense strategies are first evaluated using lightweight surrogate models. Only the most robust candidates are then deployed and tested in the DT environment to reduce computational overhead. For ease of understanding, the defense process with AED is summarized in Algorithm~\ref{alg:aed-portllm}, where defense policies are continuously refined through feedback.}

\subsection{Results and Discussion}

Fig.~\ref{losss}(a) illustrates the evolution of model accuracy under evolution attack compared with the normal Port-LLM. In the initial stage, both the normal and attacked Port-LLM exhibit similar fluctuations due to the simultaneous training and testing process. During this phase, the attack remains ineffective, as the adversary’s perturbation strategy has not yet adapted to the model’s behavior. Once training converges, the normal model maintains a stable accuracy of around 95\%, whereas the attacked model begins to show noticeable degradation across successive attack periods. From the second attack phase onward, the attacked Port-LLM shows a clear and consistent decline in accuracy compared with the normal model, indicating that the adversary progressively refines its gradient-based perturbations and increasingly compromises the model’s prediction reliability.

Fig.~\ref{losss}(b) shows the performance of the Port-LLM model equipped with the proposed AED. Compared with the normal Port-LLM, the model with AED converges more slowly during training because the defense mechanism continuously adapts to potential perturbations. \textcolor{black}{However, once convergence is achieved, the model maintains stable performance, indicating a trade-off introduced by AED: slower convergence in exchange for stronger robustness against adaptive attacks.}

\textcolor{black}{Overall, AED  offers strong robustness in dynamic attack environments because it continuously monitors system performance and updates defense policies through iterative feedback and co-evolution with the attacker. This makes AED particularly effective when adversaries adapt their strategies over time. However, AED requires additional computation and coordination overhead, which may increase latency and resource consumption during deployment. In contrast, traditional proactive defense methods, such as \cite{10944630,10726638}, are simpler, efficient to implement, and effective when attack patterns are known and stable. However, they lack flexibility once deployed and may degrade quickly when facing evolving or unseen attack strategies. Thus, AED provides long-term resilience at the cost of complexity, while proactive defenses provide efficiency but limited adaptability.}

\subsection{Takeaways}
\textcolor{black}{This case study confirms that LLM-based wireless predictors are highly sensitive to adversarial perturbations, especially when the attacker evolves over time. While proactive defenses are effective only under stable and predictable threat conditions, AED maintains robustness by continuously adjusting its defense strategies based on real-time feedback. Therefore, adaptive, co-evolving defense mechanisms are essential for securing AI-native 6G systems.}

\begin{algorithm}[ht]
\small
\caption{Port-LLM training with AED}
\label{alg:aed-portllm}
\begin{algorithmic}[1]
\Require 

Port-LLM model,
dataset $\mathcal{D}_{\text{ch}}$,
DT, LTM, thresholds $\mathcal{T}$ 

\Ensure 
Defended Port-LLM model, stable defense policy $D^*$.

\While{AED is active}
    \If{No attack is detected}
        \State PG: Generate defense set $\mathcal{D}$ and attack set $\mathcal{A}$ via LTM.
        \State FE: Train/update lightweight surrogate models to predict KPI (Port Accuracy). 
        Only top-$k$ candidates from $\mathcal{A}$ and $\mathcal{D}$ are validated on DT; others are skipped.
        \While{$\text{Accuracy} < \mathcal{T}_{\min}$}
            \State PG: Refine top-$k$ defenses by adjusting key parameters.
            \State FE: Evaluate candidates based on the Accuracy impact.
            \State CO: Perform safe deployment and real-time defense adjustments, feeding observed network conditions back into PG.
        \EndWhile
        \State CO: Save stable configuration when $\text{Accuracy} \ge \mathcal{T}_{\min}$.
        \State FE: Enhance candidates in $\mathcal{A}$ against qualified ones in $\mathcal{D}$.
        \State PG: Via LTM, generate more suitable defense.
        \State Update based on DT observations; repeat self-evolving AED.
    \Else
        \State CO: Freeze non-critical updates and limit access.
        \State FE: Identify attack type, range, and impact level.
        \State PG: Retrieve emergency defenses from LTM and DT.
        \State FE: Evaluate current defenses under the detected attacks.
        \While{$\text{Accuracy} < \mathcal{T}_{\min}$}
            \State PG: Adjust $\mathcal{D}$  for stronger protection.
            \State FE: Test the updated defenses via DT feedback.
        \EndWhile
        \If{$\text{Accuracy} \ge \mathcal{T}_{\min}$}
            \State CO: Save the stable configuration; back to normal mode.
            \State FE: Log attack traces for later analysis and retraining.
        \EndIf
    \EndIf
\EndWhile
\end{algorithmic}
\end{algorithm}

\section{Challenges and Open Issues}
Despite recent progress, securing GenAI–enabled 6G networks remains an open and evolving problem. Several challenges and open issues are summarized as follows:
\begin{itemize}
    \item The lack of standardization across vendors and platforms hinders interoperability and the exchange of trusted models, emphasizing the need for unified security protocols and cross-domain regulatory frameworks. 
    \item Most existing evaluations remain simulation-based, underscoring the need for open, large-scale testbeds and standardized benchmarks to validate adversarial attacks and defenses under realistic ISAC-6G conditions
    \item Privacy and transparency remain significant challenges as multimodal data from users and sensors are integrated into large generative models.
    \item The AED concept offers strong adaptability but requires intensive computational and energy resources. Continuous training, large-scale simulation, and real-time policy updates can quickly exceed the capacity of edge devices, especially under strict latency and power constraints in 6G networks. 
    \item With the arrival of quantum computing, attackers will be able to generate and test adversarial examples much faster than before. Quantum computing can drastically shorten the time required to break encryption, search for model weaknesses, or craft powerful perturbations, posing unprecedented risks to AI-driven communication systems.
    \item Future defenses must exploit the same computational advantage by integrating quantum-assisted training and optimization into security frameworks. Quantum parallelism can be leveraged to accelerate model verification, anomaly detection, and evolutionary policy evaluation, enabling real-time adaptation against quantum-accelerated attacks.

\end{itemize}

\section{Conclusion}
This article examined emerging security risks in GenAI-enabled 6G networks, where sensing, learning, generation, and reasoning are tightly coupled with the wireless infrastructure. We introduced a cross-domain attack taxonomy, analyzed how local compromises can cascade across ISAC, FL, DT, and LTM components, and outlined a unified defense blueprint that combines physical-layer hardening with AI-focused safeguards. Looking ahead, secure and reliable GenAI-powered 6G services will hinge on system-level co-design of security and performance, lightweight and scalable protection mechanisms, and real-time cross-layer detection and attribution. Advancing these directions together with rigorous trust verification and privacy-preserving designs for pervasive sensing and DTs will be crucial to building resilient next-generation wireless.

\bibliographystyle{IEEEtran}
\bibliography{IEEE}

\end{document}